\documentclass{iopart}

\usepackage{graphicx}

\begin{document}

\title{Nuclear physics with spherically symmetric supernova models}

\author{M Liebend\"orfer, T Fischer, C Fr\"ohlich, F-K Thielemann
and S Whitehouse}

\address{Department of Physics,
University of Basel,
Klingelbergstr. 82,
4056 Basel,
Switzerland}

\ead{matthias.liebendoerfer@unibas.ch}

\begin{abstract}
Few years ago, Boltzmann neutrino transport led to a new and reliable
generation of spherically symmetric models of stellar core
collapse and postbounce evolution. After the failure to prove the
principles of the supernova explosion mechanism, these sophisticated
models continue to illuminate the close interaction between high-density
matter under extreme conditions and the transport of leptons and energy
in general relativistically curved space-time. We emphasize that very
different input physics is likely to be relevant for the different
evolutionary phases, e.g. nuclear structure for weak rates in collapse,
the equation of state of bulk nuclear matter during bounce, multidimensional
plasma dynamics in the postbounce evolution, and neutrino cross sections
in the explosive nucleosynthesis. We illustrate the complexity of
the dynamics using preliminary 3D MHD high-resolution simulations based
on parameterized deleptonization. With established spherically symmetric
models we show that typical features of the different phases are reflected
in the predicted neutrino signal and that a consistent neutrino flux
leads to electron fractions larger than \( 0.5 \) in neutrino-driven
supernova ejecta.
\end{abstract}


\section{Introduction}

Stellar core collapse occurs towards the end of the evolution of massive
stars when their iron cores grow beyond the critical mass supportable
by the pressure of a nearly degenerate electron gas. The ensuing
gravitational collapse of the inner
core is an extreme example for the conversion of binding energy into
(neutrino-)radiation \cite{Baade.Zwicky:1934}: The transition of the
inner core of the supernova progenitor star to a compact neutron star
makes an energy of few times \( 10^{53} \) erg available for the
emission of neutrinos. This energy corresponds to a mass defect of
the remnant \( \sim 0.1 \) M\( _{\odot } \)! The first (and so far
only) detection of supernova neutrinos from SN1987A (summarised e.g.
in \cite{Bethe:1990}) recently celebrated its 20th
anniversary. The detection of supernova neutrinos provides the most
direct observational evidence for the link of a collapsing stellar
core to a supernova explosion.

Computer models with spectral neutrino
transport seem to agree on the general scenario of a delayed explosion
(see \cite{Bethe:1990} and references therein)
in which we distinguish four different phases.
During the \emph{collapse phase}, the stellar core splits into a homologously
collapsing subsonic inner core and the outer layers, which accrete with
supersonic infall velocities in the wake of the homologous collapse. The
matter has an entropy \( \sim 1 \) \( k_{B} \)/baryon and consists
of heavy, neutron-rich nuclei. Electron captures and neutrino emission
under these conditions are essential to determine the evolution of the
electron fraction through collapse and the strength of
the core rebounce in the \emph{bounce phase}, which occurs when nuclear
saturation density is exceeded at the center. Once nuclear density is
reached the low compressibility
of matter produces a stagnation wave traveling to the edge
of the homologous core, where it faces the supersonic accretion. The wave
turns into a shock which heats and dissociates the accreting matter. The
sudden appearance of free protons in a neutrino-transparent regime
leads to the emission of an
energetic neutronization burst of electron neutrinos within \( 2-5 \)
ms after bounce. This concludes the bounce phase.
The dynamic shock turns into a hydrostatically expanding
accretion front, which, throughout the ensuing \emph{accretion phase},
delimits the cold accretion flow from the hot dissociated matter piling
up on the newborn protoneutron star (PNS). The fourth phase is the
\emph{explosion phase}, where a part of the hot accumulated matter
drives the shock to larger radii into the outer layers, leading to
a supernova explosion and the ejection of matter. Another
part of the accumulated matter sinks onto the PNS or
fills the space between PNS and ejecta in the form of a neutrino-driven
wind. The explosion phase is thought to develop after an
accretion phase lasting \( 500 \) ms or more, which is of order hundred
times longer than the hydrodynamic bounce phase.

However, the obstinate difficulty
to reproduce explosions in spherically symmetric models of core collapse
and postbounce evolution
stimulated the consideration of numerous modifications and alternatives
to this basic scenario, mostly relying on multi-dimensional effects
that could not be treated in spherical symmetry. To name only a few:
It was discussed whether convection in the PNS could accelerate the
deleptonisation and increase the neutrino luminosity
\cite{Wilson.Mayle:1993}. The convective overturn between the PNS and
shock front was shown to increase the efficiency of neutrino energy
deposition \cite{Herant.Benz.ea:1994}.
Asymmetric instabilities of the standing accretion shock
\cite{Blondin.Mezzacappa.DeMarino:2003,Foglizzo.Galletti.ea:2007}
may help to push the shock to larger radii and g-mode oscillations
of the PNS may contribute to neutrino heating by the dissipation of
sound waves beween the PNS and the shock \cite{Burrows.Livne.ea:2006}.
Moreover, it has been suggested
that magnetic fields have an impact on the explosion mechanism
(see e.g. \cite{Kotake.Sato.Takahashi:2006}).
\begin{figure}
\centering\includegraphics[width=0.85\textwidth]{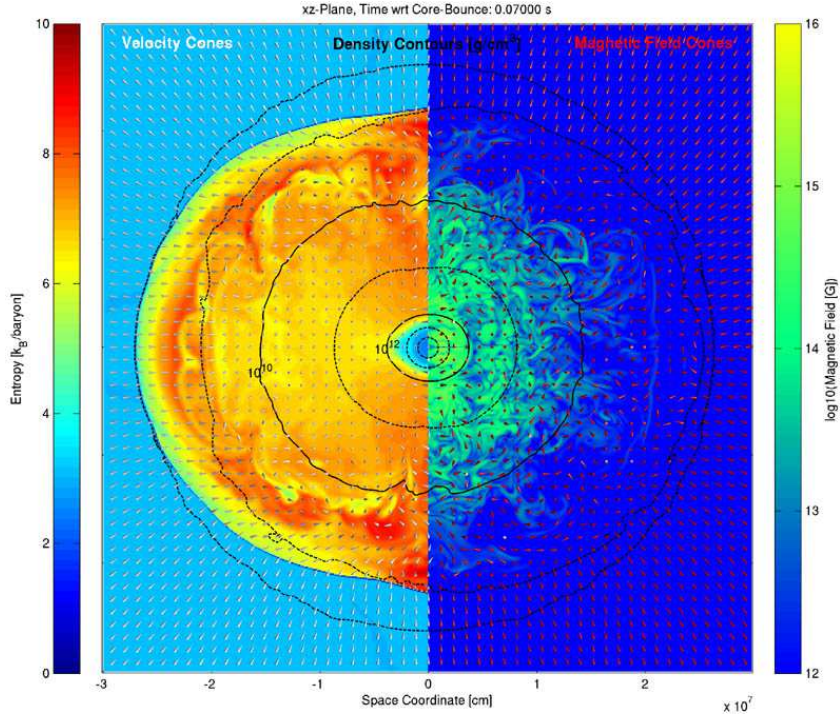}
\caption{
Illustration of the early accretion phase in a three-dimensional
simulation with parameterized neutrino physics \cite{Liebendoerfer.Pen.Thompson:2006}.
Density contours are drawn as black lines. The colour on
the left hand side indicates the specific entropy and the cones
the direction of the velocity. The colour on
the right hand side refers to the magnetic field strength and
the cones to its direction.
The cool high-density interior of the PNS and the
hot low-density accreted matter behind the standing accretion front
are clearly distinguishable.
}
\end{figure}
Most of above-mentioned modifications of the
explosion mechanism are essentially of a three-dimensional nature.
In order to illustrate the complexity of the crucial
accretion phase we show in Figure 1 a slice through a three-dimensional
simulation of core collapse and postbounce evolution of a run described
in more detail in \cite{Liebendoerfer.Pen.Thompson:2006}. Its input
physics uses the Lattimer-Swesty equation of state \cite{Lattimer.Swesty:1991}
and a parameterisation of the neutrino physics for the collapse phase
\cite{Liebendoerfer:2005}. The treatment of neutrino cooling and heating
in the postbounce phase is under development based on multi-group
diffusion, but has not yet been successfully applied in this high-resolution
run with \( 600^{3} \) zones.

Initially, spherically symmetric supernova models were the most
realistic among all feasible computer representations of the event.
With increasing observational evidence for the complexity of the
explosions (e.g. \cite{Hamuy:2003}) their primary purpose shifted from a realistic
representation to the identification and understanding of the basic
principles of the explosion mechanism. After the emergence of axisymmetric
simulations with sophisticated and computationally intensive spectral
neutrino transport
\cite{Buras.Rampp.ea:2003,Walder.Burrows.ea:2004}
spherically symmetric models
still have several assets. In the following we describe selected applications
of spherically symmetric models related to nuclear and weak interaction
input physics.

\section{Conditions obtained with GR spectral neutrino transport}

The complete general relativistic (GR)
Boltzmann equation for neutrino transport has so far only been solved
in spherically symmetric dynamical simulations
\cite{Wilson:1971,Liebendoerfer.Mezzacappa.ea:2001,Sumiyoshi.Yamada.ea:2005}.
Some of the GR effects have less effect on the explosion mechanism
(e.g. the bending of neutrino trajectories by the PNS) than others
(e.g. the GR effects included in the hydrostatic Tolman-Oppenheimer-Volkoff
equation). Hence, accurate results can be obtained by the careful
design of approximations \cite{Marek.Dimmelmeier.ea:2006}.
But spherically symmetric simulations are not only useful as a testbed
for approximations and comparisons \cite{Liebendoerfer.Rampp.ea:2005},
they can also provide a clearly arranged foundation
to determine the conditions achieved in supernova simulations
and to distinguish and
study the input physics in different regimes. Figure 2 shows an overview
of the conditions achieved in a simulation of the collapse, bounce,
and explosion of a 20 M\( _{\odot } \)
star (model B05 in \cite{Froehlich.Hauser.ea:2006}).
\begin{figure}
\centering\includegraphics[width=\textwidth]{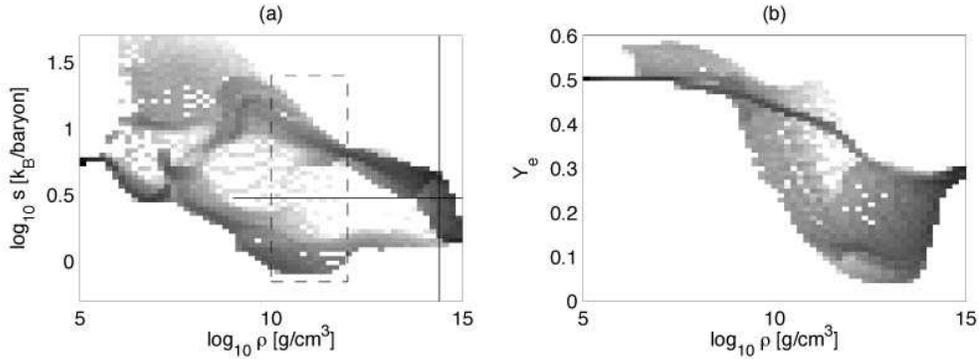}
\caption{
Overview of the conditions achieved in a simulation of the collapse, bounce,
and explosion (artificially induced) of a 20 M\( _{\odot } \) star.
Shown are two histograms of the occurence of conditions as function of
density \( \rho \), specific entropy \( s \) and electron fraction \( Y_e \).
The shading of a given bin corresponds
to \( \log _{10}\left( \int dm\, dt\right)  \) in arbitrary units,
where the integral over mass is performed over the mass \( dm \)
of matter whose thermodynamic state at a given time falls into the
bin. The integral over time extends over the duration
of a simulation. Hence, regions of dark shading correspond to states
that are assumed by considerable mass for an extended time, while
light or absent shading corresponds to conditions that are rarely assumed
in the supernova simulation. The vertical black line indicates nuclear
density. The horizontal black line indicates an entropy of \( 3 \)
k\( _{B} \)/baryon beyond which ions are dissociated. In Figure 2a,
it clearly separates the conditions of cold infalling matter on the
lower branch from the conditions of hot shocked matter on the
upper branch.
}
\end{figure}
Figure 2a shows that the four phases of the supernova scenario involve
quite distinct regimes of conditions. The conditions can be subdivided
into three different regimes: The lower dark branch reaching from
low densities up to nuclear density corresponds to cold infalling
matter containing heavy ions. Its electron fraction in Figure 2b follows
during infall the dark branch reaching from \( Y_{e}=0.5 \) to \( Y_{e}\sim 0.3 \).
The matter beyond the nuclear density threshold in Figure 2a corresponds
to bulk nuclear matter in the interior of the PNS. As one can see
from Figure 2b it does not significantly deleptonize during the simulation
because of the short mean free paths and correspondingly long diffusion times of
the neutrinos at high density. As soon as infalling matter is hit by the
accretion
shock, its entropy jumps to the upper edge of the upper branch while
its \( Y_{e} \) drops by enhanced electron capture in the dissociated
regime. The darker tinted cloud around the density of \( 10^{10} \)
g/cm\( ^{3} \) reaches entropies between \( 10 \) and \( 30 \)
k\( _{B} \)/baryon by neutrino heating between the PNS and the standing
accretion shock. The lighter tinted region toward lower densities
and higher entropies indicates conditions during matter ejection.
In Figure 2b part of the ejected
matter assumes conditions with \( Y_{e}>0.5 \) as we will discuss
later.

\section{Sensitivity of models to input physics changes}

The exchange of lepton number and energy in the models is given by
weak interaction rates between different neutrino flavours and
matter. The latter is assumed to be in thermodynamic and
nuclear statistical equilibrium except for the infalling outer layers
and expanding ejecta. The effective weak interaction rates increase
with increasing baryon density until the opacity reaches a level where
the neutrinos diffuse away more slowly than they are produced. The
rates then become Pauli-blocked. Hence there is a range of maximum
importance of the weak interaction rates at densities around the neutrinospheres
(\( \sim 10^{10}-10^{12} \) g/cm\( ^{3} \)). The intersection of
this density range (dashed box) with the two branches identified in Figure 2a leads
to two important regimes for weak interactions: The conditions on
the lower branch at \( \left( s\sim 1,Y_{e}\sim 0.35\right)  \) assumed
in infalling layers during collapse, and the conditions on the upper
branch at \( \left( s\sim 10,Y_{e}\sim 0.1\right)  \) assumed in
the hot mantle of the PNS during the accretion phase. Traditional
treatments of the weak interaction rates in supernova simulations
\cite{Bruenn:1985} recently received updates in both regimes. It was
recognized that electron captures on nuclei (and not protons) provide
the dominant deleptonization mechanism during collapse, even if the
corresponding Gamov-Teller transitions seem forbidden in the oversimplified
independent particle model of the increasingly neutron-rich nuclei
\cite{Langanke.Martinez-Pinedo.ea:2003}. The new rates lead to a smaller homologous core
and to a lower initial shock energy at bounce
\cite{Hix.Messer.ea:2003}. On the other hand, weak magnetism corrections in
the hot dissociated matter on the upper branch of Figure 2a have been
added to zeroth-order cross sections \cite{Horowitz:2002}. This results
in a slightly different balance between neutrinos and antineutrinos
in the accretion phase for all three neutrino flavours
\cite{Liebendoerfer.Mezzacappa.ea:2003}. 

The equation of state (EoS) is the second nuclear physics pillar of
supernova simulations. It has to define composition and thermodynamic
quantities as a function of density, entropy, and electron fraction
over a wide range of conditions. Firstly, the EoS can directly influence
the dynamics. A higher compressibility of matter at bounce, for example,
leads to a larger initial shock energy \cite{Bruenn:1989b}. Secondly,
the composition is another important path for the EoS to affect the simulations.
A composition with a high fraction of nuclei with large electron capture
rates will lead to faster deleptonization than a composition with
nuclei that forbid electron captures. Or, later in the accretion phase,
a large fraction of alpha particles in the heating region will decrease
the heating efficiency because they can not absorb neutrinos as
efficiently as free nucleons. Thirdly, the EoS influences the simulations
by the geometric arrangement of its consistutents. During collapse,
the ions in the low-entropy matter are spatially correlated and coherence
effects should be considered in the calculation of neutrino opacities
(\cite{Marek.Janka.ea:2005}
and references therein). The phase transition from isolated nuclei
to bulk nuclear matter is expected to involve complicated clustering
of baryons that might affect neutrino opacities as well
\cite{Watanabe.Sato.ea:2004,Horowitz.Perez-Garcia.Piekarewicz:2004,Botvina.Mishustin:2004}.

Finally, the EoS determines the simulation results by the macroscopic
structure of the PNS in hydrostatic equilibrium. Even if the high
opacity prevents the high-density regime of the PNS from affecting the
simulations directly by neutrino transport, it is the compactness
of the proto-neutron star which determines the positions of the neutrinospheres
in the gravitational potential. The positions of the neutrinospheres
have clearly visible consequences for the neutrino luminosity and
spectrum. This became evident in simulations investigating GR effects
\cite{Bruenn.DeNisco.Mezzacappa:2001,Liebendoerfer.Mezzacappa.ea:2001}
as well as in simulations comparing different EoS's
\cite{Janka.Buras.ea:2005,Sumiyoshi.Yamada.ea:2005}.

\section{Neutrino emission}

The neutrino signal during collapse and the
energetic neutronization burst few milliseconds after bounce are surprisingly
similar for progenitor stars of \( 13 \) M\( _{\odot } \) to \( 40 \)
M\( _{\odot } \). This is due to a similar size of the collapsing core
at the point of gravitational instability and to
negative feedback in the net deleptonization rate before neutrino
trapping \cite{Liebendoerfer.Mezzacappa.ea:2003}. The neutrino luminosities
during the accretion phase are more progenitor dependent. The diffusion
of neutrinos out of the cooling PNS provides the dominant contribution
to the \( \mu  \)- and \( \tau  \)-neutrino signal. An additional
contribution (of about equal size) to the electron flavour neutrino
luminosities stems from the compression of
the hot accreted matter settling on the surface
of the PNS. This contribution depends strongly on the accretion rate.

\begin{figure}
\centering\includegraphics[width=\textwidth]{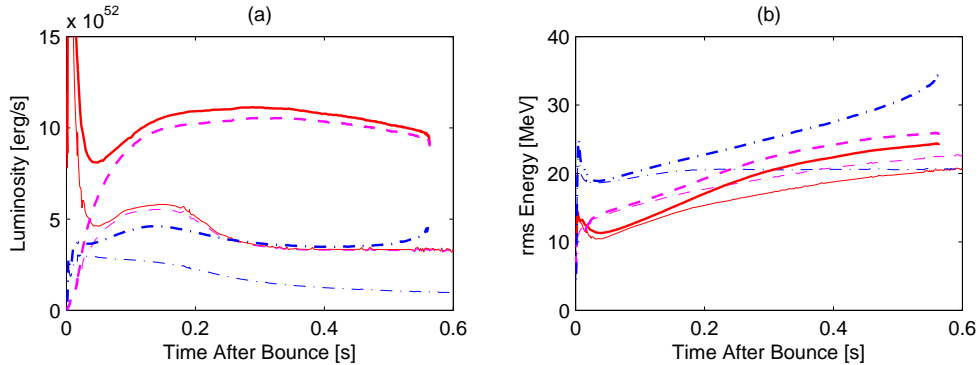}
\caption{
Shown are the neutrino luminosities and the rms energy of the neutrino
fluxes as a function of postbounce time for
two different progenitor stars: a \( 15 \)M\(_{\odot}\) star
\cite{Woosley.Weaver:1995} (thin lines) and a \( 40 \)M\(_{\odot}\) star
\cite{Umeda.pvt:2006} (thick lines). The solid, dashed and dash-dotted
lines refer to
electron neutrinos, electron antineutrinos and heavy neutrinos respectively.
The values are given at \( 200 \) km radius in the frame comoving
with the fluid.
}
\end{figure}
Figure 3a shows the time evolution
of the neutrino luminosities for two different progenitor stars after
the decay of the neutronization burst. The thin lines correspond to
an extension of model G15 \cite{Liebendoerfer.Rampp.ea:2005}. While
the luminosities of the heavy neutrinos (dash-dotted line) show a progressively
cooling PNS, the electron flavor neutrinos (solid and dashed lines)
reveal a significant drop in the accretion rate around \( 200 \)
ms after bounce before they level off to an extended stationary accretion
phase. This behaviour is expected for a PNS that does not compress
significantly because it is quite far from reaching its maximum stable
mass. Since \( \mu \)- and \( \tau \)-neutrino producing layers in the
PNS cool faster than they are compressionally heated, the rms energies of
the heavy neutrinos fall below the rms energies of the electron flavour
neutrinos, which stem from the continued accretion of hot matter.
The accretion rate would suddenly drop at the time of an eventual
explosion (not obtained in this run). The electron flavour neutrino
luminosities would then suddenly decay and asymptote to the level
of the heavy neutrino luminosities of the cooling PNS.

A different
scenario occurs for a \( 40 \) M\( _{\odot } \) progenitor star
\cite{Umeda.pvt:2006} whose luminosities in the postbounce evolution
are also displayed in Figure 3a (thick lines).
We also see the decreasing accretion rate reflected in the electron
flavour neutrino emission (solid and dashed lines), but the heavy
neutrino luminosities go through a minimum and raise again (dash-dotted
line). This is due to the hydrostatic compactification of the PNS
as it approaches the maximum stable mass determined by the high-density
EoS. Hot accreted matter is compressed to densities where the main
\( \mu  \)- and \( \tau  \)-neutrino production takes place so that
their luminosities increase
\cite{Liebendoerfer.Messer.ea:2004,Fischer.Liebendoerfer.Mezzacappa:2007,Sumiyoshi.Yamada.Suzuki:2007}.
We want to emphasize with this comparison that large differences in
the neutrino luminosities (and spectra) in models for different explosion
scenarios and launched from different progenitor models do not just
classify as 'uncertainties', they reflect various dynamical processes
at the center of a collapsed star, like the exact time of core-bounce,
the evolution of the accretion rate, the compressibility of the PNS,
the formation of a black hole, or the onset of the supernova explosion.

Before the neutrino signal reaches a terrestrial observer, one has
to account for neutrino flavour transformations. Of special interest
for the neutrino heating after the onset of the explosion are possible
collective modes \cite{Duan.Fuller.Qian:2006,Fogli.Lisi.ea:2007}.

\section{Spherical ejection models and nucleosynthesis}

Supernova nucleosynthesis predictions traditionally rely on artificially
induced explosions, replacing the central engine either with a
parameterized kinetic energy piston or a thermal bomb.
The explosion energy and the placement of the mass cut (separating
ejected matter from matter which is assumed to fall back onto the
neutron star) are tuned to recover the observed explosion energy and
ejected \( ^{56} \)Ni mass. Both approaches are largely compatible
\cite{Aufderheide.Baron.ea:1991} and justifiable for the outer stellar
regions, but most of the Fe-group nuclei are produced in the inner
regions which are most affected by the details of the explosion mechanism.
The electron fraction \( Y_{e} \) is an indispensable quantity for
the description of the explosive nucleosynthesis in the innermost
ejecta. It is set by weak interactions in the explosively burning
layers, i.e.\ electron and positron capture, beta-decays, and neutrino
or anti-neutrino captures.

We examined the effects of both electron and neutrino captures in
the context of spherically symmetric simulations with Boltzmann neutrino
transport \cite{Froehlich.Hauser.ea:2006}. In order to provoke the
ejection of matter in spherically symmetric models,
we had to modify the simulations in the accretion
phase. Having PNS convection \cite{Wilson.Mayle:1993} or convective
turnover \cite{Herant.Benz.ea:1994} in mind, we artificially enhanced
the neutrino emission from the PNS or, alternatively, the neutrino
absorption efficiency in the hot mantle surrounding it. Both approaches
serve as matter ejection models with a consistently emerging mass
cut. Similar simulations using tracer particles from two-dimensional
simulations \cite{Buras.Rampp.ea:2003} have been performed in
\cite{Pruet.Woosley.ea:2005}. Also in this case, artificial adjustments
to the simulations were needed to remedy the failure of the underlying
models to produce self-consistent explosions. Also in both cases,
the neutrino transport could not be run to later times and the simulations
were mapped to a simpler model to continue the simulation. Despite these shortcomings,
these simulations reveal the significant impact of neutrino
interactions on the composition of the ejecta. At early times, the
inner ejecta are electron-degenerate and electron capture dominates.
Due to neutrino heating in the heating region and the expansion
this degeneracy is lifted and electron-emitting
neutrino absorption reactions start to dominate the change of \( Y_{e} \).
Eventually the electron chemical potential drops below the mass difference
between the neutron and proton. At this energy scale, the proton is
favored because of its slightly larger binding energy. We found that
all our simulations that lead to an explosion by neutrino heating
developed a proton-rich environment around the mass cut with \( Y_{e}>0.5 \)
\cite{Froehlich.Hauser.ea:2006}.

The nucleosynthesis in proton-rich ejecta has been investigated in
\cite{Thielemann.Hauser.ea:2002,Umeda.Nomoto:2005,Pruet.Woosley.ea:2005}.
The global effect of the proton-richness is the removal of previously
documented overabundances of neutron rich iron peak nuclei \cite{Woosley.Weaver:1995,Thielemann.Nomoto.Hashimoto:1996}.
Production of \( ^{58,62} \)Ni is suppressed while \( ^{45} \)Sc
and ~\( ^{49} \)Ti are enhanced. The results for the elemental abundances
of scandium, cobalt, copper, and zinc are closer to those obtained by
observation
\cite{Froehlich.Hauser.ea:2006}. However, the neutrino
interactions are not only responsible for the proton-richness of the
environment, they also transform protons into neutrons by antineutrino
capture so that (n,p)-reactions substitute for the slow \( \beta  \)-decays
in the waiting point nuclei, allowing significant flow to \( A>64 \)
by the \( \nu \)p-process \cite{Froehlich.Martinez-Pinedo.ea:2006}.
This process turns out to have a significant impact on the nucleosynthesis
in the early neutrino wind \cite{Pruet.Hoffman.ea:2006,Wanajo:2006}.

\section*{Acknowledgements}

This work was supported in part by grants No. PP002-106627/1,
No. 200020-105328/1 and No. IB7320-110996/1
from the Swiss National Science Foundation and
the EU program ILIAS N6 ENTApP WP1.

\section*{References}


\end{document}